\documentclass[reprint,raggedbottom,aps,prl,amsfonts,amsmath,amssymb,floatfix]{revtex4-1}

\usepackage{graphicx,hyperref} 
\usepackage[all]{hypcap} 
\usepackage{xcolor}
\definecolor{red}{HTML}{CC0000}
\definecolor{blue}{HTML}{306EFF}
\definecolor{darkgray}{HTML}{242220}
\color{darkgray}
\hypersetup{colorlinks=true,linkcolor=blue,citecolor=blue,urlcolor=blue}
\setcitestyle{super} 

\renewcommand{\v}[1]{\mathbf{#1}}
\newcommand{\Ang}[0]{\, \mathring{\mathrm{A}}}
\newcommand{\eV}[0]{\, \mathrm{eV}}
\newcommand{\kB}[0]{k_\t{B}}
\newcommand{\K}[0]{\, \t{K}}

\renewcommand{\t}[1]{\text{#1}}

\begin{document}

\onecolumngrid
  \begin{center}
    \Large \textbf{\color{red} This manuscript has been published in an open-access format in Nature Communications, 
    DOI: \href{https://doi.org/10.1038/s41467-020-16892-4}{\texttt{10.1038/s41467-020-16892-4}}}
  \end{center}
\twocolumngrid

\makeatletter
\def\bstctlcite{\@ifnextchar[{\@bstctlcite}{\@bstctlcite[@auxout]}}
\def\@bstctlcite[#1]#2{\@bsphack
  \@for\@citeb:=#2\do{\edef\@citeb{\expandafter\@firstofone\@citeb}
  \if@filesw\immediate\write\csname #1\endcsname{\string\citation{\@citeb}}\fi}
  \@esphack}
\makeatother
\bstctlcite{fix_citation_style}

\title{Uncovering the effects of interface-induced ordering of liquid on crystal growth using Machine Learning}
\author{Rodrigo Freitas}
\email{freitas@stanford.edu}
\author{Evan J. Reed}
\affiliation{Department of Materials Science and Engineering, Stanford University, Stanford, CA 94305, USA}
\date{\today}

\begin{abstract}
  \begin{center}
    \textbf{ABSTRACT}
  \end{center}

  The process of crystallization is often understood in terms of the fundamental microstructural elements of the crystallite being formed, such as surface orientation or the presence of defects. Considerably less is known about the role of the liquid structure on the kinetics of crystal growth. Here atomistic simulations and Machine Learning methods are employed together to demonstrate that the liquid adjacent to solid-liquid interfaces presents significant structural ordering, which effectively reduces the mobility of atoms and slows down the crystallization kinetics. Through detailed studies of silicon and copper we discover that the extent to which liquid mobility is affected by interface-induced ordering (IIO) varies greatly with the degree of ordering and nature of the adjacent interface. Physical mechanisms behind the IIO anisotropy are explained and it is demonstrated that incorporation of this effect on a physically-motivated crystal growth model enables the quantitative prediction of the growth rate temperature dependence.
\end{abstract}
\maketitle

\section{Introduction}
Crystallization from the melt (Fig.\ref{fig:MD_simulation}) is a pervasive process in industry, from metal casting for structural applications to the Czochralski process for semiconductor wafer growth for electronics. It is important to control and understand the crystal growth process because it is at this stage that the material's microstructure morphology is created, which in turn defines the material's properties. Consequently, a great deal of effort has been put into understanding the complex interplay between structure, thermodynamics, and kinetics that governs the process of crystal growth\cite{asta_review,libbrecht_review,bcf}. This has led to a mechanism-based understanding of crystallization\cite{chernov_book,pimpinelli_book,saito_book} in terms of the microstructural elements of the crystallite being formed. For example, the character of the solid surface in contact with the liquid is known to affect the growth rate, with atomically rough surfaces leading to faster growth rates than flat low-index surfaces and their vicinals. Considerably less attention has been put in understanding the effects that the liquid adjacent to the solid-liquid interface has on the process of crystal growth.

Atomic events leading to crystal growth are thermally activated processes taking place in the free-energy landscape illustrated in Fig.~\ref{fig:Wilson-Frenkel}a. The rate of crystallization is proportional to $\exp(-\beta \Delta E_\t{a})$ while the melting rate is proportional to $\exp[-\beta (\Delta E_\t{a} +\Delta \mu)]$, where $\Delta E_\t{a}$ is the activation energy for solidification, $\Delta \mu$ is the difference in chemical potential between the liquid and solid phases, $\beta^{-1} \equiv \kB T$, and $\kB$ is the Boltzmann constant. The balance of these two rates results in the following equation for the overall growth rate:
\begin{equation}
  \label{eq:wf}
  r(T) = k(T) \Big\{ 1 - \exp\big[-\beta \Delta \mu(T)\big] \Big\},
\end{equation}
where $k(T) \equiv k_0 \exp(-\beta \Delta E_\t{a})$ is known as the kinetic factor. In this model, known as the Wilson-Frenkel\cite{wilson,frenkel} (WF) model, the activation energy for solidification is taken as the energy barrier for diffusion in the liquid, $\Delta E_\t{a} = \Delta E_\t{d}$, because crystallizing atoms must undergo the same self-diffusion process that occurs in the associated liquid phase. It is often found that the WF cannot quantitatively predict results from simulations or experiments\cite{si_wf_exp,harrowell_nat_mat}. This notorious discrepancy, while largely unsolved, has been attributed to changes in mobility of the supercooled liquid in the vicinity of the crystal interface that would cause $\Delta E_\t{a} > \Delta E_\t{d}$, but no physical mechanism has been demonstrated to explain the origin of this effect.

Here we employ atomistic simulations and Machine Learning (ML) together to show that the solid-liquid interface induces partial ordering of the nearby liquid during crystal growth. Our approach is successfully applied to two different families of materials: semiconductors and metals. We find that the interface-induced ordering (IIO) affects the mobility of liquid atoms and thus slows down the crystal growth kinetics. The physical mechanism behind the IIO is explained and we demonstrate that by accounting for this effect it is possible to derive predictive models for crystal growth. 

\section{Results}

\begin{figure*}
  \centering
  \includegraphics[width=1.00\textwidth]{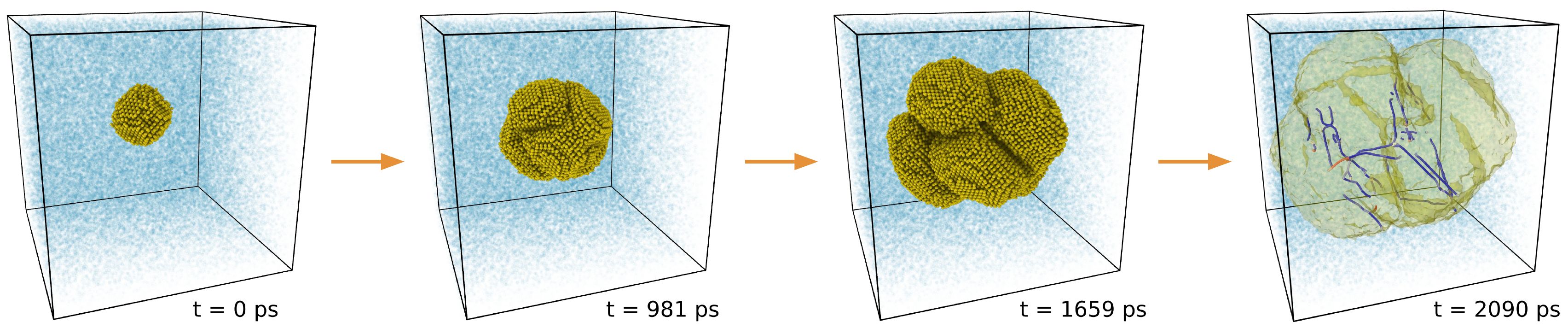}
  \caption{\label{fig:MD_simulation} \textbf{Crystal growth simulations.} Snapshots of a crystal growth simulation of silicon using Molecular Dynamics. The system initially contains a small crystalline seed (yellow atoms) surrounded by liquid (transparent blue atoms). Shown in the last frame is the dislocation network\cite{DXA_1,dxa,ovito} formed during the growth process (edge dislocations are colored blue while screw dislocations are shown in red). See Supplementary Video 1 for the complete video and Supplementary Video 3 for the dislocation dynamics.} 
\end{figure*}

\begin{figure*}
  \centering
  \includegraphics[width=1.00\textwidth]{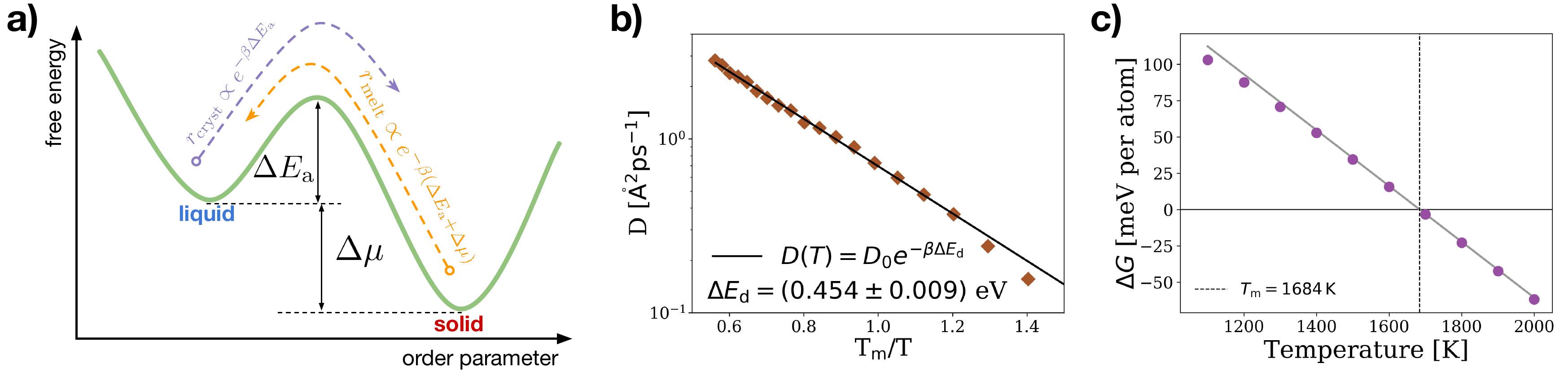}
  \caption{\label{fig:Wilson-Frenkel} \textbf{Wilson-Frenkel model.} \textbf{a)} Free-energy landscape for solidification according to the Wilson-Frenkel model. The activation energy for solidification is given by $\Delta E_\t{a}$, while $\Delta \mu$ is the chemical potential difference between the liquid and solid. Atomic events contributing to crystal growth are thermally activated processes occurring with rate $r_\t{cryst}$ for atoms moving from the liquid to the solid and $r_\t{melt}$ for atoms moving from the solid to the liquid. \textbf{b)} Arrhenius plot of the liquid diffusivity as a function of temperature. The solid black line is the result of a least-squares fit to the data for $T \ge T_\t{m}$. The energy barrier for diffusion in the liquid $\Delta E_\t{d}$ is used as input parameter in the Wilson-Frenkel model. \textbf{c)} Difference in free energy $\Delta G$ between the solid and liquid phases. The thermodynamic melting temperature occurs at $\Delta G(T_\t{m}) = 0$. The solid gray line is a guide to the eye to accentuate deviations from the linear behavior.} 
\end{figure*}

\begin{figure*}
  \centering
  \includegraphics[width=1.00\textwidth]{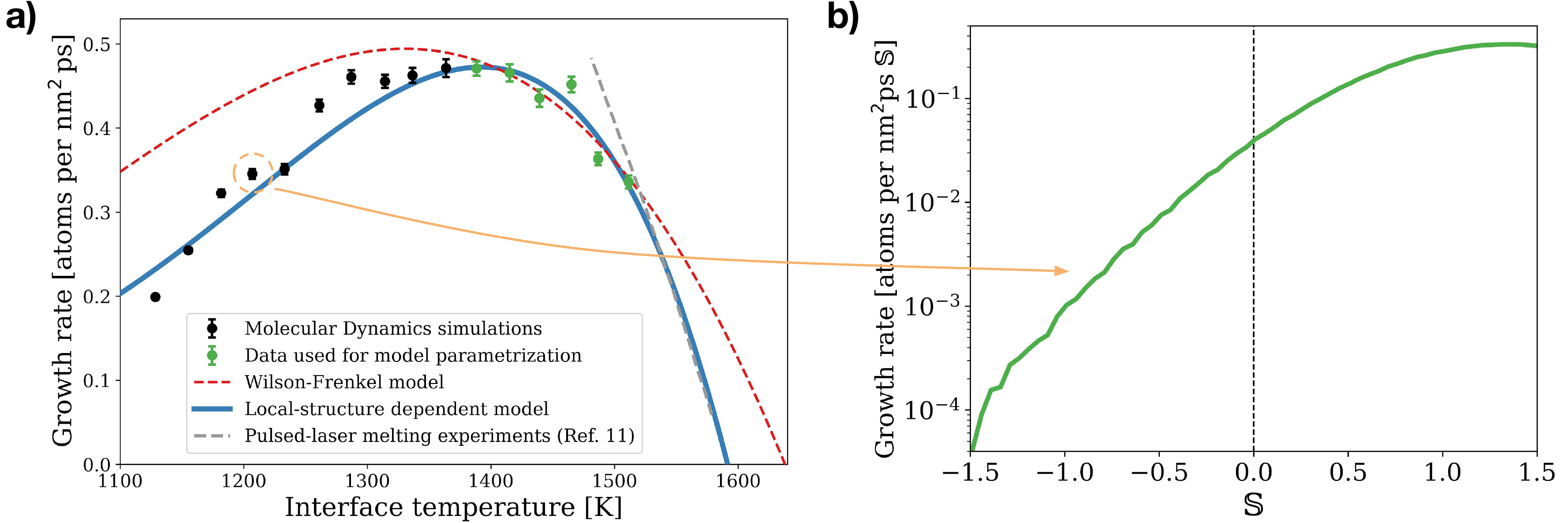}
  \caption{\label{fig:growth_rate} \textbf{Local-structure dependent crystal growth model.} \textbf{a)} Crystal growth rate of silicon versus interface temperature. Both models (Wilson-Frenkel and local-structure dependent model) were parametrized using only the data from simulations at $T \ge 1388\K$. Notice that the growth rate is given per unit of effective area of the total crystallite (see Supplementary Note 4) and the error bar represents the 95\% confidence interval around the mean (see Methods). Data for laser-pulsed melting experiments was extracted from reference \citenum{si_epitaxial}. \textbf{b)} Growth rate at $T \approx 1233\K$ decomposed as a function of the local structure, as encoded by $\mathbb{S}$, surrounding crystallizing atoms. Notice that the growth rate varies almost four orders of magnitude as $\mathbb{S}$ changes.}
\end{figure*}

\textbf{Crystal growth simulations.} We performed Molecular Dynamics simulations of crystalline silicon growth from its melt employing a simulation geometry akin to laboratory experiments of crystal growth: a crystalline seed is introduced in the liquid and its growth is monitored over the course of the simulation (see Fig.~\ref{fig:MD_simulation} and Supplementary Video 1). This setup allows the different microstructural elements of the growing crystallite to interact naturally (see Supplementary Videos 2 and 3), as they would in a crystal growth experiment. For this geometry $\Delta \mu = \Delta G - \kappa \gamma / \rho_\t{s}$, where $\Delta G$ is the difference in free energy between the liquid and the crystal, and the second term is due to the Gibbs-Thomson effect, with $\rho_\t{s}$ being the density of the solid, $\gamma$ the interfacial free energy, and $\kappa = 2 / \mathcal{R}_\t{eff}$ is a geometrical factor where $\mathcal{R}_\t{eff}$ is the effective crystallite radius. All the above parameters of the WF model were computed (Figs.~\ref{fig:Wilson-Frenkel}b and \ref{fig:Wilson-Frenkel}c) in order to compare the model predictions against simulation results (for calculation details see Methods and Supplementary Note 5 and 6). The comparison between model and simulations is shown in Fig.~\ref{fig:growth_rate}a, where it is evident that the WF model does not predict the growth rate for temperatures it was not fitted to. \\

\textbf{Machine Learning encoding of atomic crystallization events.} Historically, simpler simulation geometries have been favored as a way to isolate certain microstructural elements, which are then probed separately\cite{hoyt_prb,harrowell_nat_mat,steps_prb}. Our use of the geometry shown in Fig.~\ref{fig:MD_simulation} makes the simulation more physically relevant at the expense of greatly diminishing the amount of information that can be inferred due to the lack of a crystal growth model that accounts for all microstructural elements present and their respective interactions. Moreover, it also becomes challenging to decipher the atomic events at play due to the sheer complexity of the environment that atoms are embedded in. Here these obstacles are overcome by employing ML algorithms to systematically encode and classify the structure surrounding liquid atoms during crystallization events. Our approach builds on recently-proposed ML strategies for the construction of a structural quantity (namely softness $\mathbb{S}$) that captures the propensity for atomic rearrangements to occur in disordered atomic environments, such as in glasses\cite{dogus_prl,softness} and inside grain boundaries\cite{softness_gb}.

\begin{figure*}
  \centering
  \includegraphics[width=1.00\textwidth]{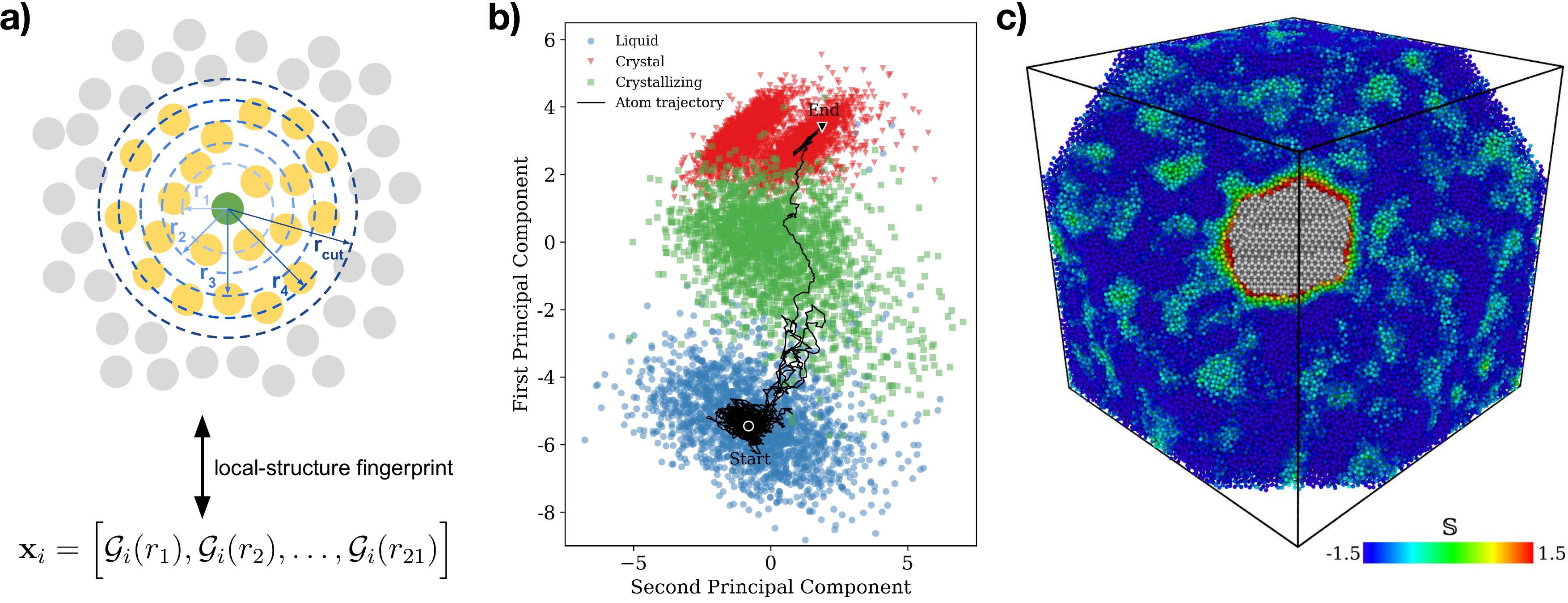}
  \caption{\label{fig:ML_encoding} \textbf{Machine Learning encoding of atomic crystallization events.} \textbf{a)} The local structure (atoms in yellow) surrounding a central atom (green) is encoded using a set of 21 radial structure functions $\mathcal{G}(r_n)$, each evaluated at a different radial distance $r_n$ from the central atom, with $n = 1, 2, 3, \ldots, 21$. Together these functions comprise the atom's local-structure fingerprint, denoted as $\v{x}_i$ for the $i$th atom. \textbf{b)} Atomic trajectory during crystallization as encoded by the time-evolution of $\v{x}_i\in\mathbb{R}^{21}$. The $\mathbb{R}^{21}$-space was represented in two dimensions using the first two components of the Principal Component Analysis method. \textbf{c)} Cross section of a snapshot of the initial stages of silicon growth. Liquid atoms are colored according to their softness ($\mathbb{S}$) value, while atoms in the crystalline phase are colored in gray. The clusters of liquid atoms far from the crystallite with $\mathbb{S} \approx 0$ are due to dynamical heterogeneities in the supercooled liquid.}
\end{figure*}

The structural characterization of local atomic environments is realized by assigning to each atom $i$ a local-structure fingerprint $\v{x}_i$ constructed from a set of 21 radial structure functions\cite{parrinello_prl,dogus_prl} $\mathcal{G}(r)$, as illustrated in Fig.~\ref{fig:ML_encoding}a. Furthermore, atoms are labeled into three possible categories according to their first-neighbor's arrangement: liquid and crystal atoms have arrangement patterns statistically identical to the bulk liquid and bulk crystal, respectively. Meanwhile, crystallizing atoms have arrangement patterns intermediary between the other two labels (see Methods and Supplementary Note 2 for more details). It is possible to observe how these three groups of atoms are spread in the $\mathbb{R}^{21}$-space of local-structure fingerprints $\v{x}_i$ with the help of an algorithm known as Principal Component Analysis. With this method, a dimensionality reduction transformation is performed to create a $\mathbb{R}^2$ representation of the $\mathbb{R}^{21}$ data, as shown in Fig.~\ref{fig:ML_encoding}b. Superimposed in this figure is also the trajectory of an atom that undergoes crystallization over the course of the simulation.

Atoms assume varied local-structure fingerprints $\v{x}_i$ depending on both the surrounding liquid structure and the nearby interface morphology. In order to quantify these variations in microstructure we proceed as follows. First, an ML algorithm known as Support Vector Machine\cite{svm,sklearn,libsvm} is employed to find the hyperplane that optimally separates the crystallizing atoms from the liquid atoms in the $\mathbb{R}^{21}$-space of $\v{x}_i$. Then, the distance  of each atom $i$ from the hyperplane ($\mathbb{S}_i$, known as softness\cite{dogus_prl,softness,softness_gb}) is measured: atoms with $\mathbb{S}_i > 0$ lie on the crystallizing side of the hyperplane, while $\mathbb{S}_i < 0$ atoms lie on the liquid side. This approach is found to correctly classify liquid and crystallizing atoms with an accuracy of 96\%.  It is important to realize that $\mathbb{S}$ is not an order parameter because it was not designed to track the change from the liquid to the solid phase. Instead, $\mathbb{S}$ measures the propensity of an atom in the liquid phase to undergo the process of crystallization.

Shown in Fig.~\ref{fig:ML_encoding}c is a simulation snapshot with atoms colored according to their softness value (see also Supplementary Videos 4 and 5). In this figure $\mathbb{S}$ is seen to capture the structural signs of dynamical heterogeneity in the supercooled liquid far from the crystal, with clear indications of strong spatial correlations. These fluctuating heterogeneities have recently been shown to be preferential sites for crystal nucleation\cite{water_nucleation,valeria_nucleation}. Thus, Fig.~\ref{fig:ML_encoding}c establishes that $\mathbb{S}$ is indeed capable of capturing subtle signs of structural ordering in liquids. \\

\textbf{Local-structure dependent crystal growth model.} It is possible now to decompose the crystal growth rate of Fig.~\ref{fig:growth_rate}a as a function of the local structure using $\mathbb{S}$. For example, in Fig.~\ref{fig:growth_rate}b it is shown that the total growth rate at $T \approx 1233\K$ varies by almost four orders of magnitude as the local structure changes. For this reason, we propose to address the limitations of the WF model by taking into account the local structure surrounding the crystallizing atoms through an explicit dependence on $\mathbb{S}$:
\begin{equation}
  \label{eq:rf}
  r(T,\mathbb{S}) = k(T,\mathbb{S}) \Big\{ 1 - \exp\big[-\beta \Delta \mu(T,\mathbb{S})\big]\Big\},
\end{equation}
with $\Delta \mu(T,\mathbb{S}) = \Delta G(T) - \kappa \gamma(\mathbb{S})/\rho_\t{s}$. Indeed, accounting for the information about the local structure contained in plots such as Fig.~\ref{fig:growth_rate}b results in a crystal growth model with predictive capabilities, as shown in Fig.~\ref{fig:growth_rate}a (see Supplementary Note 3 and 4 for details on the model calculation). Notice how the local-structure dependent (LSD) model is capable of predicting the growth rate for a wide range of temperatures (i.e. $T < 1388\K$) not included in the model parametrization. In particular, the experimentally measured growth rate and its slope show much better agreement with our LSD model, Eq.~\eqref{eq:rf}, than with the WF model, Eq.~\eqref{eq:wf}. In the Supplementary Note 3 we show that the variables introduced by the dependence on $\mathbb{S}$ are not independent parameters. Thus, the improved reproduction and prediction of simulation results cannot be attributed simply to Eq.\eqref{eq:rf} exhibiting higher capacity or flexibility in modeling complex relationships when compared to Eq.\eqref{eq:wf}.

\begin{figure*}
  \centering
  \includegraphics[width=1.00\textwidth]{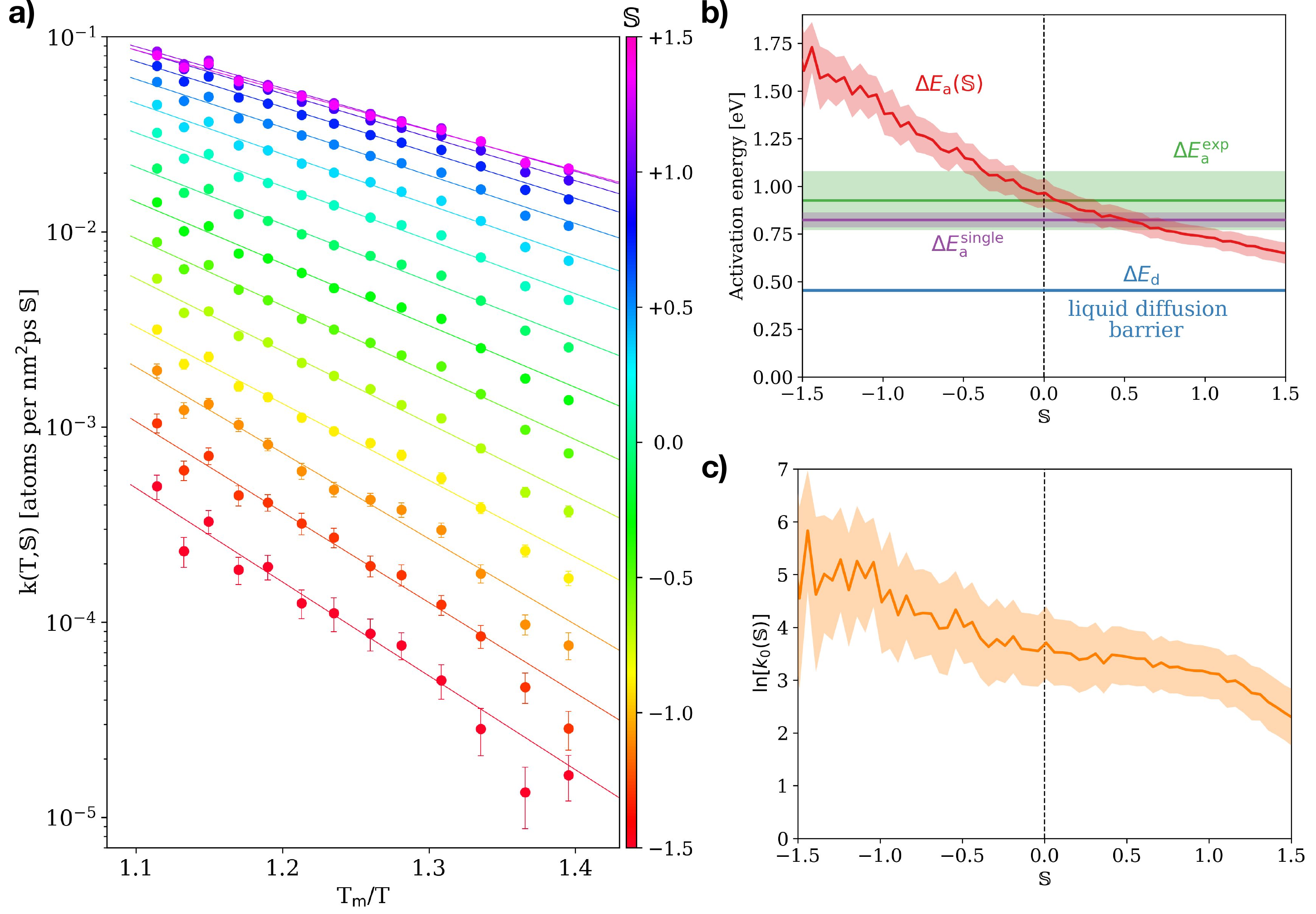}
  \caption{\label{fig:LSD_parameters} \textbf{Parameters of the local-structure dependent crystal growth model.} \textbf{a)} Arrhenius plot of the crystallization kinetic factor for different local structures in silicon, as encoded by softness ($\mathbb{S}$). The Arrhenius dependence on temperature suggests that each value of $\mathbb{S}$ can be interpreted as a thermally activated and independent channel for crystallization with a well-defined energy scale. Notice that the glass transition temperature, $T_\t{m} / T_\t{g} \approx 1.68$\cite{si_glass}, is beyond the temperature range of the figure. \textbf{b)} Dependence of the activation energy barrier for solidification on the local structure. The extent to which the liquid properties are affected by interface-induced ordering seems to vary greatly, from a negligible change ($\Delta E_\t{a} \approx \Delta E_\t{d}$) to an impressive variation of over $1\,\t{eV}$. The experimental results ($\Delta E_\t{a}^\t{exp}$) were obtained from reference \citenum{si_wf_exp} and assume a single activation energy, i.e. Eq.\eqref{eq:wf} with $\Delta E_\mathrm{a}$ as a free parameter for fitting. For the purpose of comparison we computed the equivalent quantity in our simulations: $\Delta E_\t{a}^\t{single}$, which is shown to agree with the experimental results within the accuracy of the error bars. \textbf{c)} Arrhenius prefactor dependence on $\mathbb{S}$. The decrease of three orders of magnitude with $\mathbb{S}$ implies that there are less rearrangement pathways leading liquid atoms to the activated crystallizing state as $\mathbb{S}$ increases. All error bars reported are the standard deviation of their respective parameters, except for $\Delta E^\t{exp}_\t{a}$ where we report the value in reference \citenum{si_wf_exp}.}
\end{figure*}

We now turn to investigate the ramifications of the LSD model $r(T,\mathbb{S})$ and uncover the source of its predictive capabilities. The kinetic factor $k(T,\mathbb{S})$, shown in Fig.~\ref{fig:LSD_parameters}a, is observed to be a strong function of the local structure, varying by as much as three orders of magnitude with $\mathbb{S}$. For each value of $\mathbb{S}$ the kinetic factor shows an Arrhenius-like temperature dependence $k(T,\mathbb{S}) = k_0(\mathbb{S}) \exp[-\beta \Delta E_\t{a}(\mathbb{S})]$. This striking outcome suggests that each value of $\mathbb{S}$ corresponds to a thermally-activated and independent crystallization channel with well-characterized energy scale. Such a picture is reminiscent of our traditional understanding of crystallization in terms of the solid-liquid interface morphology, with different values of $\mathbb{S}$ encoding the influence of different microstructural elements. But here $\mathbb{S}$ encodes more than just the crystal local microstructure: it also encodes the variation in the structure of the liquid. The variation of liquid properties with local structure is reflected in the dependence of the activation energy barrier of these crystallization channels with $\mathbb{S}$ shown in Fig.~\ref{fig:LSD_parameters}b, where it can be seen that $\Delta E_\t{a}(\mathbb{S})$ varies over $1\,\t{eV}$ with $\mathbb{S}$. Additionally, the activation energy decreases monotonically with $\mathbb{S}$ and seems to approach the energy barrier for diffusion $\Delta E_\t{d}$ (Fig.~\ref{fig:Wilson-Frenkel}b) asymptotically. Hence, the mobility of liquid atoms close to the solid-liquid interface seems to vary greatly, from a negligible change ($\Delta E_\t{a} \approx \Delta E_\t{d}$) compared to bulk liquid to a dramatic reduction in mobility due to the increase in $\Delta E_\t{a}$. This change in the liquid structure due to the presence of the solid-liquid interface is known as IIO, Supplementary Figure 5b shows that the structural change does indeed lead to local ordering.

Varying $\mathbb{S}$ also has a pronounced effect on the Arrhenius prefactor $k_0(\mathbb{S})$, as indicated in Fig.~\ref{fig:LSD_parameters}c, which decreases by three orders of magnitude with $\mathbb{S}$. Because $\ln[k_0(\mathbb{S})]$ can be interpreted as the product of an entropic contribution\cite{softness_gb} to the free energy barrier and a term involving the population of crystallizing atoms with softness $\mathbb{S}$, the observed decrease in the prefactor indicates that there are less rearrangement pathways leading liquid atoms to the activated state (i.e. to crystallization) as $\mathbb{S}$ increases. Hence, Figs.~\ref{fig:LSD_parameters}b and \ref{fig:LSD_parameters}c together indicate that from all observed local-structure arrangements surrounding crystallizing atoms, only very few lead to low energy barriers. Additionally, Fig.~\ref{fig:growth_rate}b indicates that these few channels with low energy barriers are the ones contributing the most to the overall growth rate.

Next, we examine how the free energy of the solid-liquid interface to which atoms attach varies with $\mathbb{S}$, which should give us a glimpse of the microstructure at the crystallite surface. Figure \ref{fig:gamma}a shows that $\gamma(\mathbb{S})$ decreases monotonically with $\mathbb{S}$, starting at large values -- corresponding to high-index interfaces -- and reaching interfacial free energy values characteristic of low-index interfaces in silicon. This finding implies that the decrease in Arrhenius prefactor $k_0(\mathbb{S})$ with softness (Fig.~\ref{fig:LSD_parameters}c) leads to fewer rearrangement pathways because crystallization events with large positive values of $\mathbb{S}$ happen at low-index surfaces and their vicinals, which naturally offer less crystallization sites than high-index interfaces. Despite the scarcity of crystallization sites offered by low-index interfaces and their vicinals, Fig.~\ref{fig:growth_rate}b shows that they contribute the most to the overall growth rate, with $70\%$ of all atoms attaching to interfaces with $\mathbb{S} \ge 0.75$. This observation is confirmed by direct measurement of the distribution of crystal surfaces to which crystallizing atoms attach: Fig.~\ref{fig:gamma}b reveals strong preferential attachment to a wide variety of $(111)$ vicinals. The high-intensity spot around $(435)$ corresponds to step-step separation distances from $15\Ang$ to $24\Ang$ (Fig.~\ref{fig:gamma}c), indicating that the majority of the crystallization events take place on vicinal surfaces with well-separated steps, which is exactly what is expected for silicon\cite{si_review}. Figure \ref{fig:gamma}b also shows a smaller amount of events occurring at high-index interfaces, further validating the above observations. Notice in Supplementary Video 1 that the crystallite also exhibits signs of rough interfaces, thus it is possible that a small fraction of the identified high-index are actually rough.

\textbf{Applicability to a different family of materials.} In order to verify that our approach in creating LSD predictive models of crystal growth is not particular to silicon (or semiconductors) we apply it in the development of a crystal growth model for an elemental metal, namely copper (see Methods and Supplementary Note 7 for simulation details). The resulting model is shown in Fig.~\ref{fig:copper}a, where it can be seen that the LSD model of copper also correctly predicts the growth rate at temperatures at which it was not parametrized on (i.e. it is a predictive model), while the WF model is not capable of reproducing simulation results at temperatures to which it was not fitted, similarly to what was observed for silicon in Fig.~\ref{fig:growth_rate}a. Moreover, analysis of the parameters of the LSD model of copper (Fig.~\ref{fig:copper}b) shows that all parameters present the same trend with $\mathbb{S}$ as observed for silicon, including the Arrhenius behavior for the kinetic factor (shown in Supplementary Figure 19c). 

The major difference observed between the LSD models of silicon and copper is that $k_0(\mathbb{S})$ and $\Delta E_\t{a}(\mathbb{S})$ assume much larger values for copper. We attribute this to the predominance of rough interfaces in metallic systems. In contrast to semiconductors, interfaces in metallic systems typically do not advance by the lateral motion of steps. Instead, metal interfaces often advance by atomic attachment directly on top of them, leading to growth normal to the interface itself. This growth mechanism is reflected in Fig.~\ref{fig:copper}a (inset), where it is seen that atomic attachments occur directly on $(001)$ and $(111)$ interfaces -- leading to normal growth -- instead of vicinals of $(111)$ as observed for silicon in Fig.~\ref{fig:gamma}b (compare also Fig.~\ref{fig:MD_simulation} to Supplementary Figure 14 and Supplementary Video 1 to Supplementary Video 7). Normal growth leads to the formation of atomically rough interfaces that offer a much larger amount of atomic disorder than well-structured high-index interfaces. Hence, rough interfaces present a larger availability of sites for liquid atoms to attach, leading to much higher values for $k_0(\mathbb{S})$ due to the numerous atomic pathways leading to crystallization. The predominance of rough interfaces also explains the larger values of $\Delta E_\t{a}(\mathbb{S})$ observed for copper, but this explanation will be postponed until the Discussion section, where the connection will be discussed in the light of the effects of IIO on crystal growth.

\section{Discussion}
Solid-liquid interfaces in equilibrium are known to affect the structure of the nearby liquid by imparting some amount of order on it\cite{iio_structure,iio_kinetics,iio_siau,iio_review,iio_1,iio_2,iio_3,layering_1,layering_2}. Here, we have established that the IIO of the liquid also occurs during the process of crystal growth -- a dynamic situation in which the solid-liquid interface is not in equilibrium. The observed IIO seems to decrease the mobility of liquid atoms through changes in the activation barrier for crystallization $\Delta E_\t{a}$, Fig.~\ref{fig:LSD_parameters}b, effectively slowing down the crystallization kinetics.

\begin{figure}
  \centering
  \includegraphics[width=0.45\textwidth]{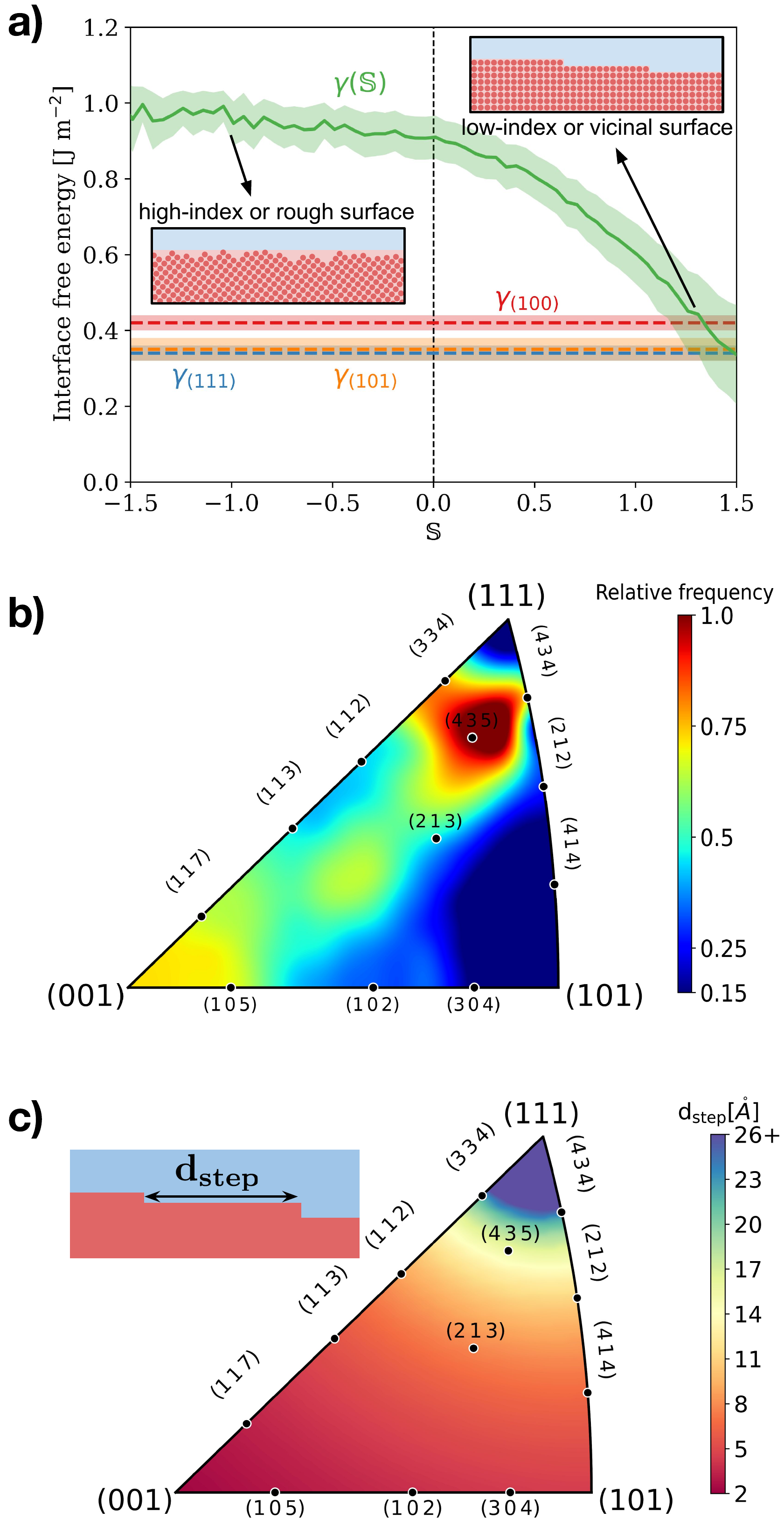}
  \caption{\label{fig:gamma} \textbf{Interface structure effect on crystal growth.} \textbf{a)} Free energy of the solid-liquid interface to which atoms with softness $\mathbb{S}$ attach. Large positive values of $\mathbb{S}$ have interface free energies characteristic of low-index interfaces in silicon, while negative values of $\mathbb{S}$ are found to be associated with high-index interfaces. Interfacial free energies $\gamma_{(100)}$, $\gamma_{(101)}$, and $\gamma_{(111)}$ and their respective error bars were obtained from reference \citenum{gamma_si}. The error bar of $\gamma(\mathbb{S})$ is the standard deviation. \textbf{b)} Distribution of interfaces to which the crystallizing atoms attach, showing a strong preference for $(111)$ vicinals and a smaller amount of crystallization events on high-index interfaces. \textbf{c)} Step-step separation distances ($d_\t{step}$) for steps on $(111)$ surfaces. Interfaces for which $d_\t{step}$ is much larger than the interatomic distance are vicinals (i.e. composed of $(111)$ facets well separated by steps), while interfaces with $d_\t{step}$ of the order of the interatomic distance are high-index interfaces in which individual steps cannot be discerned anymore.}
\end{figure}

\begin{figure*}
  \centering
  \includegraphics[width=1.000\textwidth]{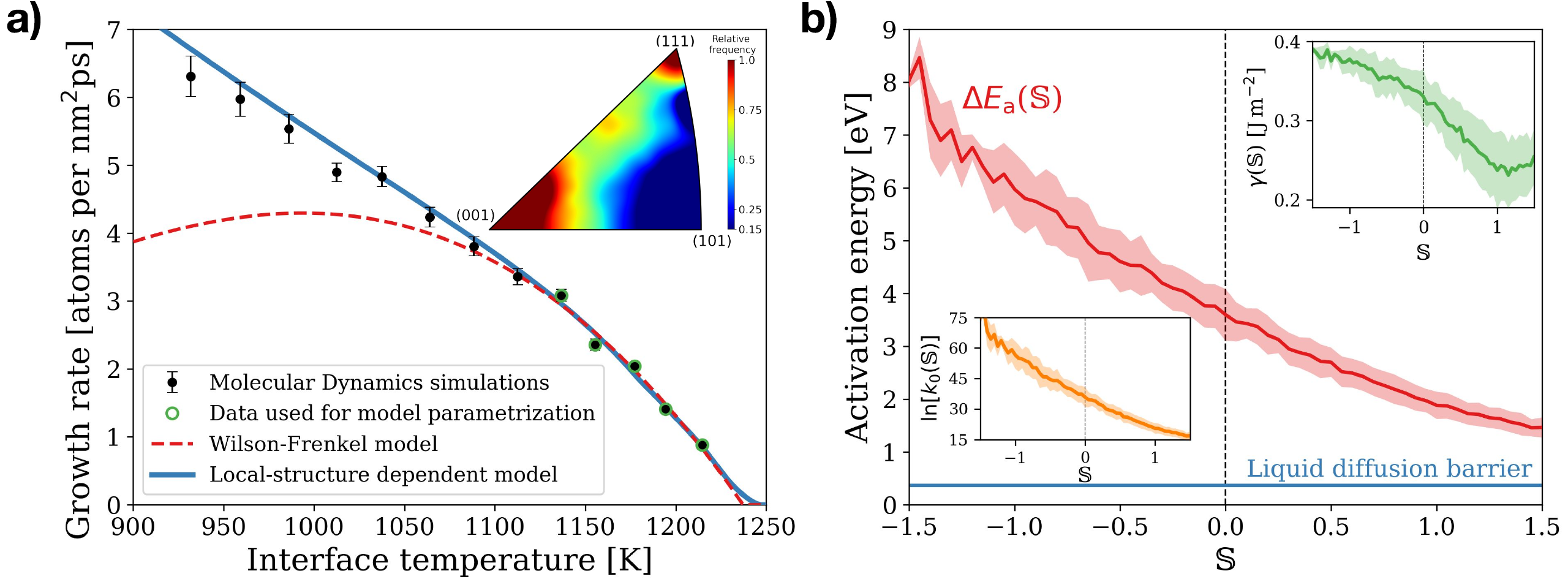}
  \caption{\label{fig:copper} \textbf{Local-structure dependent model of crystal growth for copper.} \textbf{a)} Crystal growth rate versus interface temperature for copper. Both models (Wilson-Frenkel and local-structure dependent model) were parametrized using only the data from simulations at $T \ge 1137\K$. The error bars represent the 95\% confidence interval around the mean (see Methods). \textbf{b)} Dependence of the activation energy for solidification on softness ($\mathbb{S}$). The inset shows the surface free energy and the Arrhenius prefactor. The interface-induced ordering of the liquid affects the process of crystal growth of metals in the same manner as it was observed for silicon. The major difference when compared to the results for silicon is that $k_0(\mathbb{S})$ and $\Delta E_\t{a}(\mathbb{S})$ assume much larger values due to the predominance of rough interfaces in metallic systems. Error bars are standard deviations.}
\end{figure*}

\begin{figure}
  \centering
  \includegraphics[width=0.45\textwidth]{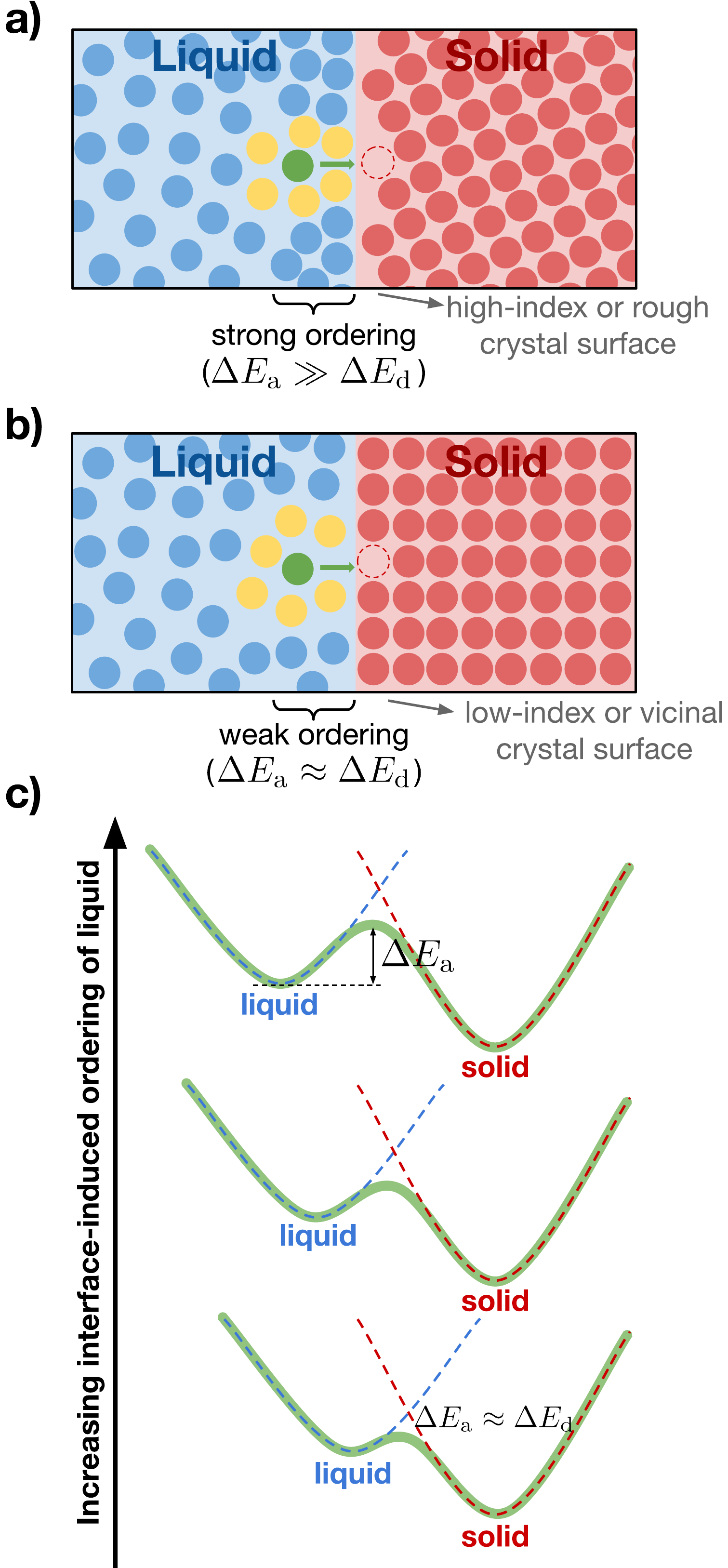}
  \caption{\label{fig:mechanism} \textbf{Interface-induced ordering mechanism.} \textbf{a)} Illustration of how interface-induced ordering of the liquid alters the local structure around crystallizing atoms and affects the activation energy for solidification. The crystallizing atom (green) has its local structure (illustrated here only by its first neighbors) affected by the nearby solid-liquid interface. This effect is anisotropic: high-index or rough interfaces interact strongly with the liquid and cause significant ordering of the liquid, which becomes rigid, resulting in large activation energies $\Delta E_\t{a}$ when compared to the barrier for diffusion in the liquid $\Delta E_\t{d}$. \textbf{b)} Low-index interfaces interact weakly with the liquid and cause very small ordering, resulting in low $\Delta E_\t{a}$. \textbf{c)} Schematic illustration of the effect of interface-induced ordering of the liquid on the free-energy landscape of crystallization. Note how the liquid free-energy basin moves to the left with increasing ordering, causing $\Delta E_\mathrm{a}$ to become progressively and continuously larger.}
\end{figure}

Comparison of Figs.~\ref{fig:LSD_parameters}b and \ref{fig:gamma}a reveals that the IIO of the liquid is anisotropic, i.e. it depends on the interface orientation and microscopic levels of roughness. The trend (illustrated in Fig.~\ref{fig:mechanism}) is such that low-index surfaces and their vicinals (corresponding to large positive $\mathbb{S}$ in Fig.~\ref{fig:gamma}a) cause weak ordering, resulting in smaller activation energies (i.e. $\Delta E_\t{a}(\mathbb{S})$ close to the energy barrier for atomic diffusion in the liquid bulk $\Delta E_\t{d}$). Meanwhile, high-index interfaces (negative $\mathbb{S}$ in Fig.~\ref{fig:gamma}a) cause strong ordering of the liquid, which becomes rigid and results in activation energies much larger than $\Delta E_\t{d}$. However, even in the case of strong IIO the activation energy ($\approx1.75\eV$) is still much smaller than the $\approx4.6\eV$\cite{cSi_diffusion} barrier for vacancy-mediated self-diffusion in crystalline silicon. This indicates that the structural order of the liquid affected by IIO is nowhere as substantial as crystalline order.

The physical cause of the IIO anisotropy is that the interaction between the crystal surface and the liquid is mediated by the amount of dangling bonds on the crystal surface. Thus, strong liquid ordering (and slower mobility) is observed at high-index interfaces because these interfaces interact more strongly with the liquid since they present more dangling bonds when compared to low-index interfaces and its vicinals. This mechanism is illustrated in Figs.~\ref{fig:mechanism}a and ~\ref{fig:mechanism}b, while its effect on the free-energy landscape of the system is illustrated schematically in Fig.~\ref{fig:mechanism}c. Notice that this mechanism also explains why copper has much larger values of $\Delta E_\t{a}(\mathbb{S})$ (Fig.~\ref{fig:copper}b) while having similar energy barrier for diffusion in the liquid $\Delta E_\t{d}$: rough interfaces are predominant in copper and these interfaces have much stronger interactions with the liquid when compared to low-index and vicinal surfaces, which are predominant in silicon.

Dynamical heterogeneities present in the liquid (Fig.~\ref{fig:ML_encoding}c) also affect the coordination of atoms\cite{water_nucleation,water_mobility,si_ll}. For this reason, it is reasonable to expect that they contribute to the $\approx\!1\eV$ dispersion in $\Delta E_\t{a}(\mathbb{S})$ observed in Fig.~\ref{fig:LSD_parameters}b. Nonetheless, there is no evident reason to believe that their effect is anisotropic since dynamical heterogeneities have origin in random thermal fluctuations.

In conclusion, we have discovered that the IIO of liquids strongly affects the process of crystal growth in metals and semiconductors. It is found that the modified structure of the liquid nearby solid-liquid interfaces reduces the mobility of liquid atoms, an effect shown to be essential in order to build a predictive model of the growth rate temperature dependence. Indeed, the construction of such predictive model was only possible by identifying and incorporating in the model the family of all thermally-activated events -- each with its own energy scale -- leading liquid atoms to the crystal phase. Our work elevates the liquid structure to the same level of importance as the crystal surface morphology in understanding crystallization, a knowledge that can enable material advances through the incorporation of liquid-structure engineering as a novel pathway for synthesis. Our results were only made possible by employing atomistic simulations and ML together. The strength of this combined approach is that one can perform complex simulations and yet glean physical insight from notoriously haphazard atomic environments. This innovative application of ML in materials science blends conventional scientific methods with data science tools to produce physically-consistent predictive models and novel conceptual knowledge.


\clearpage
\onecolumngrid
\begin{center}
  \textbf{\LARGE Methods}
\end{center}
\vspace{0.4cm}
\twocolumngrid

\vspace{0.4cm}
\noindent \textbf{Silicon crystal growth simulations.} The Molecular Dynamics (MD) simulations were performed using the Large-scale Atomic/Molecular Massively Parallel Simulator (LAMMPS\cite{lammps}) software, with the interactions between silicon atoms described by the Stillinger-Weber\cite{stillinger_weber} interatomic potential. The timestep was selected as approximately $1/56\t{th}$ of the period of the highest-frequency phonon mode of this system, or $\Delta t = 1 \, \t{fs}$. The crystal growth simulations contained $500,000$ atoms and were initialized with a spherical crystalline seed of approximately $3,000$ atoms in the diamond cubic structure. The lattice parameter for the atoms in the crystal seed was chosen taking into account dilation due to thermal expansion, then the remainder of the simulation cell was filled with randomly distributed atoms at the equilibrium liquid density for that temperature at zero pressure. 

The system was equilibrated by first relaxing the liquid atoms using a Conjugate Gradient\cite{bulatov_book} algorithm for $200$ steps. Next, the liquid was equilibrated at finite temperature using the Bussi-Donadio-Parrinello\cite{bdp} (BDP) thermostat for $3\,\t{ps}$ with a damping parameter of $0.1\,\t{ps}$. Finally, liquid atoms were equilibrated for $2\,\t{ps}$ at zero pressure and finite temperature using the same thermostat just described and a chain Nos\'e-Hoover barostat\cite{barostat_1,barostat_2,barostat_3,barostat_4,tuckerman_book} with damping parameter of $1\,\t{ps}$ and a chain length of three, allowing only for isotropic dilation/contraction of the system. During the entirety of this equilibration process the crystalline seed atoms were kept frozen at their equilibrium crystal structure with fixed lattice parameters (i.e. they did not dilate/contract with the liquid atoms). After equilibration the BDP thermostat and chain Nos\'e-Hoover barostat were applied to the entire system, both with damping parameter of $1.0\,\t{ps}$, to maintain the system at finite temperature and zero pressure for a total of $3\,\t{ns}$ during which snapshots were recorded every $1\,\t{ps}$. The crystal growth process can be seen in the Supplementary Video 1. Snapshots saved from the MD simulations were subsequently relaxed using $20$ steps of the Steepest-Descent\cite{bulatov_book} algorithm. The crystal growth simulations were performed at temperatures ranging from $1125\K$ to $1500\K$ in intervals of $25\K$.

The damping parameter for the thermostat was selected conservatively such that the liquid diffusivity was not affected by the presence of the thermostat, i.e. it had the same value within the statistical uncertainty as the diffusivity computed without a thermostat. Thus, the thermostating of the crystal growth simulation was performed gently as to not affect the kinetics of the system. See Supplementary Note 6 for more details on the diffusivity calculations.

\vspace{0.4cm}
\noindent \textbf{Phase identification.} In order to identify to which phase (liquid or crystal) each particle belongs, we used the order parameter introduced by \citealt*{alpha}. The complete description and analysis of the construction of this order parameter can be found in the Supplementary Note 2. Ultimately, this method provides us with a parameter $\alpha_i(t)$ for each atom $i$ at the time $t$ of each MD snapshot. The physical interpretation of this parameter is that $\alpha_i$ is the fraction of bonds that atom $i$ makes that resemble bonds in a perfect crystal structure. As shown in the Supplementary Note 2, the parameter $\alpha_i$ correctly identifies atoms in the perfect crystal or bulk liquid with accuracy of $100\%$ within the statistical uncertainty. It is important to notice that although $\alpha_i$ can discern between liquid and crystal atoms, it does not differentiate between crystalline structures. We confirm that the silicon atoms are indeed crystallizing in the diamond cubic structure by performing the Polyhedral Template Matching\cite{ptm,ovito} analysis.

\vspace{0.4cm}
\noindent \textbf{Encoding atomic dynamics (ML labeling).} The dynamics of each atom was encoded using the time evolution of the $\alpha_i(t)$ order parameter. A representative plot\footnote{Unless otherwise noted, all figures were produced using \texttt{Matplotlib}\cite{matplotlib}.} of $\alpha_i(t)$ is shown in Supplementary Figures 4b and 4c. Notice that due to thermal fluctuations the instantaneous value of $\alpha_i(t)$ for atoms in the liquid and crystal phases might differ from their perfect values of $0.0$ and $1.0$ respectively, even after the short Steepest-Descent relaxation. Hence, we perform a moving-window average of $\alpha_i(t)$ with window length of $20\,\t{ps}$ and use the window-averaged $\bar{\alpha}_i(t)$ to label the atomic dynamics as illustrated in Supplementary Figure 4a. Atoms with $\bar{\alpha}_i(t) = 0$ for $t\in[t_0-\tau_\ell,t_0+\tau_\ell]$ receive label $y_i = -1$ at time $t_0$. These are atoms deep in the liquid phase that will not be transitioning to the crystal state in the near future, neither have tried to transition in the near past. From the analysis of curves such as in Supplementary Figure 4c we choose $\tau_\ell = 15\,\t{ps}$ as a reasonable value. Next we identify atoms that have just started to move out of the bulk liquid (i.e. crystallizing atoms) as those within a $20\,\t{ps}$ window from the point where $\bar{\alpha}_i = 0.25$, i.e. $y_i = 1$ for $t \in [t_0-\tau, t_0+\tau]$ where $\bar{\alpha}_i(t_0) = 0.25$ and $\tau = 10\,\t{ps}$. See Supplementary Note 2 for more details on the labeling process.

\vspace{0.4cm}
\noindent \textbf{Local-structure fingerprint (ML features).} The local structure surrounding each atom was characterized using a set of radial structure functions\cite{parrinello_prl,dogus_prl}:
\[
  \mathcal{G}_i(r) = \sum_{j=1}^{n(i)} \exp\Big[-(r_{ij}-r)^2/2\sigma^2\Big],
\]
where $i$ is the atom whose local structure is being described, $n(i)$ is the number neighbors of $i$ within a cutoff radius $r_\t{cut}$, $r_{ij}$ is the distance between atom $i$ and one of its neighbors $j$, $r$ and $\sigma$ are two parameters that define the radial structure function. These smoothly varying functions of $r$ count the number of neighbors of $i$ at a distance $r$. In this interpretation, parameter $r$ represents the radial distance from $i$ at which we are counting the number of neighbors, while $\sigma$ adjusts how smoothly the function varies as atoms move in and out of the distance $r$ vicinity. We have used a grid-search\cite{dl_book} algorithm to perform the hyperparameters tuning (see Supplementary Note 1 for more details), resulting in $\sigma = 0.5\Ang$, $r_\t{cut} = 10.8\Ang$, and $r_n = (2.0 + 0.4n)\Ang$, with $n = 0, 1, 2, \ldots, 20$. With this set of $21$ radial structure functions -- one for each value of $r$ -- the local-structure fingerprint of each atom $i$ was built as a vector:
\[
  \v{x}_i = \Big[ \mathcal{G}_i(r_1), \mathcal{G}_i(r_2), \ldots, \mathcal{G}_i(r_{21}) \Big].
\]

\vspace{0.4cm}
\noindent \textbf{Softness calculation.} The data was assembled by pairing the dynamic labels $y_i$ with their corresponding structural fingerprint $\v{x}_i$. Then, $10,000$ $(y_i,\v{x}_i)$ pairs were randomly selected and equally divided between the $y = -1$ and $y = 1$ classes to train a Support Vector Machine\cite{svm,sklearn,libsvm} (SVM) classifier. The SVM algorithm finds the hyperplane of the form $\v{w} \cdot \v{x} - b = 0$ that optimally separates the two classes, where $\v{w}$ and $b$ are the parameters that define this hyperplane. Before training the SVM classifier the elements of the fingerprints were standardized\cite{dl_book} to have zero mean and standard deviation of one. The optimal hyperplane found, denoted by the parameters $\v{w}^*$ and $b^*$, correctly separates the two classes with an accuracy of 96\%. See Supplementary Note 1 for more details about how these parameters are found and an in-depth analysis of the quality of the classifier computed. All results shown here were obtained using data from the crystal growth simulation at $T = 1500\K$ to train the SVM classifier. However, the results can be reproduced within the statistical uncertainty when training at any other temperature, as shown in the Supplementary Note 3.

Once the SVM classifier has been trained it was applied to the entire data set, composed of $27.5$ million data points (excluding the data used for training, hyperparameter tuning, and cross validation). The value of softness\cite{softness} for each data point (or atom) is the signed distance from the hyperplane, or $\mathbb{S}_i = \v{w}^* \cdot \v{x}_i - b^*$ for each atom $i$.

\vspace{0.4cm}
\noindent \textbf{Parameter estimation and temperature extrapolation.} In order to test how predictive the LSD and WF models are we performed the model parameterization of both models using only the data collected for low undercooling (i.e. $T \ge 1388\K$) and observed how well the model predicts the temperature dependence for higher undercoolings (i.e. temperatures as low as $1128\K$). As shown in Fig.~\ref{fig:growth_rate}a the LSD model is capable of predicting the growth rate at temperatures it was not parametrized on, while the WF model only reproduces the simulation results at temperatures it was fitted to. We attribute this to the fact that the LSD model accounts for the Arrhenius family of thermally activated atomic events leading to crystal growth, as labeled by $\mathbb{S}$. This is fundamental physical information that is not incorporated in the WF model.

For Figs.~\ref{fig:LSD_parameters} and \ref{fig:gamma}a only, the $\mathbb{S}$ dependence of the parameters of the LSD model (i.e. $\Delta E_\t{a}$, $k_0$, and $\gamma$) was measured after reparametrizing this model using the data from all simulations with $T \ge 1206\K$. The reparametrization was necessary only in order to reduce the statistical uncertainty of the parameters measured. This range of temperatures was chosen because below $T = 1206\K$ the kinetic factor $k(T,\mathbb{S})$ showed signs of non-Arrhenius behavior for some values of $\mathbb{S}$ due to the approaching glass transition temperature. Notice that below $T = 1206\K$ the bulk liquid diffusivity also shows signs of departure from the Arrhenius behavior. Thus, there are no reasons to expect the Arrhenius behavior for $k(T,\mathbb{S})$ to hold below $T = 1206\K$ because the mobility of liquid atoms close to the crystal surface is smaller than in the bulk due to IIO effects.

\vspace{0.4cm}
\noindent \textbf{Principal Component Analysis.} Figure \ref{fig:ML_encoding}b was obtained by applying the Principal Component Analysis\cite{dl_book} (PCA) to a data set containing equal amounts of crystallizing, liquid, and crystal, for a total of $60,000$ data points. Crystal atoms were defined as those for which $\bar{\alpha}_i(t) = 1.0$ (bulk crystal atoms) or $\bar{\alpha}_i(t) = 0.75$ (stacking fault atoms) for $t \in [t_0-\tau_\t{x}, t_0+\tau_\t{x}]$ with $\tau_\t{x} = 15\,\t{ps}$. From the PCA we obtained the components of each data point along the two eigenvectors with largest eigenvalues, which are used to plot Fig.~\ref{fig:ML_encoding}b. The atom trajectory was obtained by applying the same PCA transformation along a single $3\,\t{ns}$ trajectory of an atom in a simulation at $1500\K$.

\vspace{0.4cm}
\noindent \textbf{Growth rate determination.} The number of atoms in the crystallite $N(t)$ at any given time $t$ was determined as the number of atoms with $\bar{\alpha}_i(t) > 0.25$. From this information the effective crystallite radius $\mathcal{R}_\t{eff}(t)$ (shown in Supplementary Figure 10a) was computed assuming a spherical shape (which results in $\kappa = 2 / \mathcal{R}_\t{eff}$, where $\kappa$ is the geometrical factor in the Gibbs-Thomson term). The growth rate for each temperature, shown in Fig.~\ref{fig:growth_rate}a, was determined by a linear fit of $N(t)$ over the time interval for which $\mathcal{R}_\t{eff} \in [80\Ang,100\Ang]$. This interval is such that the crystallite is small enough to not be affected by finite-size effects, but large enough to give the system time to equilibrate into a steady-state growth condition. See Supplementary Note 4 for a more detailed analysis.

Error bars in Fig.~\ref{fig:growth_rate}a represent the $95\%$ confidence interval as computed using the bootstrap method with $1000$ samples of the same size as the original distribution.

\vspace{0.4cm}
\noindent \textbf{Interface temperature.} When studying crystal growth, it is important to differentiate between the temperature of the supercooled liquid surrounding the crystal (but far from the interface) from the solid-liquid interface temperature. Under steady-growth conditions these two temperatures will differ because of the latent heat released at the interface and the finite rate of heat transport. Here, the interface temperature was computed by considering only the kinetic energy of atoms with $\bar{\alpha}_i \in (0.15,0.75)$. This interval of $\bar{\alpha}$ was selected because it includes both, interfacial liquid and interfacial crystal atoms. Supplementary Figure 10b shows the interface temperature under steady-state growth as a function of the surrounding liquid bath temperature. See Supplementary Note 4 for more details.

\vspace{0.4cm}
\noindent \textbf{Crystal surface analysis.} Figure \ref{fig:gamma}b was obtained by constructing a polyhedral surface mesh around the crystallite (i.e. atoms with $\bar{\alpha}_i(t) > 0.25$) using the algorithms in reference \citenum{surface_mesh} (as implemented in Ovito\cite{ovito}) with a probe-sphere radius of $3.0\Ang$ and a smoothing level of $10$. From this mesh the surface directions were inferred and averaged over the time interval for which the crystal growth occurs in a steady state. The data for constructing Fig.~\ref{fig:gamma}b was obtained by finding the orientation of the closest surface to each crystallizing atom.

The only atomically smooth surfaces in silicon are $\{111\}$ surfaces\cite{si_review}, consequently steps can only exist in these surfaces. Hence, the step-step separation distance shown in Fig.~\ref{fig:gamma}c was computed assuming that $(111)$ faceting occurs at all surface orientations.

\vspace{0.4cm}
\noindent \textbf{Solid and liquid free energies.} The accurate calculation of the solid and liquid free energies is crucial in crystal growth studies. As shown in the Supplementary Note 5, employing approximations such as the quasi-harmonic approximation results in the underestimation of the predicted growth rates by as much as $36\%$. For this reason, we performed the solid and liquid free energies calculations using state-of-the-art nonequilibrium thermodynamic integration methods that make no approximating assumptions on the physical characteristics of the system. The crystal free energy was determined using the nonequilibrium Frenkel-Ladd\cite{fl,fl_2,cms_freitas} (FL) and the Reversible Scaling\cite{rs,cms_freitas} (RS) methods, following closely the approach described in reference \citenum{cms_freitas}. For both methods a system of $21,952$ silicon atoms in the diamond cubic structure was employed. The thermodynamic switching was performed in $200\,\t{ps}$ for each direction, before which the system was equilibrated for $20\,\t{ps}$. The FL switching was realized for temperatures ranging from $100\K$ to $2000\K$ in intervals of $100\K$. For each temperature the switching was repeated in five independent simulations to estimate the statistical uncertainty. Similarly, the RS switching was also repeated five times. The $S$-shaped function was employed in the FL switching, while the RS switching was performed with $T_\t{i} = 100\K$ and $T_\t{f} = 2000\K$ under the constant $\t{d} T/\t{d}t$ constraint. The system's center-of-mass was kept fixed for the FL and RS simulations, while a Langevin\cite{langevin_thermostat,comp_sim_liquids} thermostat with damping parameter of $0.1\,\t{ps}$ was applied. For the RS method a chain Nos\'e-Hoover barostat with damping parameter of $1\,\t{ps}$ and chain length of three was used to keep zero pressure. The absolute free energies and a comparison with the harmonic and quasi-harmonic approximations\cite{brent_fultz_book} can be seen in Supplementary Figure 11.

Liquid free energies were computed using the Uhlenbeck-Ford\cite{uf,cms_uf} (UF) and RS\cite{cms_uf,rs} methods, following closely the approach described in reference \citenum{cms_uf}. The liquid free energy calculations had the same number of atoms, switching time, equilibration time, and thermostat as the crystal free-energy calculations. The liquid density was the equilibrium density at zero pressure, with the thermal expansion dilation taken into account. For the UF method we used $p=50$, $\sigma = 1.5\Ang$, and a cutoff radius of $r_\t{c} = 5\sigma$. The UF switching was performed linearly with time, while the RS switching had the same time dependence as the crystal with $T_{i} = 2000\K$ and  $T_\t{f} = 1100\K$ (the lower final temperature $T_\t{f}$ was chosen to avoid the liquid vitrification at low temperatures). For both methods -- UF and RS -- the switchings were repeated in five independent simulations to estimate the statistical uncertainty.

\vspace{0.4cm}
\noindent\textbf{Copper.} All results for copper were obtained from simulations that followed the exact same specifications as the simulations described above for silicon. The only modifications performed are described in this session.

The interaction between copper atoms was described using the embedded-atom method\cite{eam} interatomic potential of \citet{eam_cu}. The timestep was selected as approximately $1/66\t{th}$ of the period of the highest-frequency phonon mode of this system, or $\Delta t = 2 \, \t{fs}$. The crystal growth simulations contained $1,000,000$ atoms (notice that this is twice the size of the simulations for silicon) and were initialized with a spherical crystalline seed of approximately $24,000$ atoms (eight times larger than for silicon) in the face-centered cubic structure. The difference in system size allowed us to explore much lower undercoolings for the crystal growth simulations, which ran for a total of $2\,\t{ns}$ per temperature. In contrast to the simulations for silicon, the snapshots saved from MD simulations were not subsequently relaxed before computing structural parameters because it has been shown that energy minimizations lead to significant crystallization in metallic systems\cite{ultrafast}. The crystal growth simulations were carried out at temperatures ranging from $900\K$ to $1200\K$ in intervals of $25\K$. The growth rate for each temperature, shown in Fig.~\ref{fig:copper}a, was determined by a linear fit of $N(t)$ over the time interval for which $\mathcal{R}_\t{eff} \in [100\Ang,120\Ang]$, as detailed in Supplementary Note 4.

The local-structure fingerprint $\v{x}_i$ for the copper atoms was composed of a set of 37 radial structure functions defined by the following parameters: $\sigma = 0.3\Ang$, $r_\t{cut} = 20.6\Ang$, and $r_n = (2.0 + 0.5n)\Ang$, with $n = 0, 1, 2, \ldots, 37$. These parameters were obtained through the same hyperparameter optimization process applied to silicon, as described in the Supplementary Note 1. The SVM classifier trained with the data collected for copper had accuracy of $97\%$. The results presented here were obtained using data from the simulation at $T = 1200\K$ to train the SVM classifier. The total size of the data set for copper was 77.9 million data points.

All results for the LSD model for copper (including Fig.~\ref{fig:copper} and Supplementary Figure 19c) were obtained using data from simulations at $T \ge 1136\K$. Below this temperature the kinetic factor showed signs of non-Arrhenius behavior for some values of $\mathbb{S}$ due to the approaching glass transition temperature, similarly to what was observed for silicon. Notice that below $T = 1136\K$ the bulk liquid diffusivity also shows signs of departure from the Arrhenius behavior. Thus, there are no reasons to expect the Arrhenius behavior for $k(T,\mathbb{S})$ to hold below $T = 1136\K$ because the mobility of liquid atoms close to the crystal surface is smaller than in the bulk due to IIO effects.

The solid and liquid free energies were computed for systems containing 19,652 copper atoms. The FL and RS methods were applied with a switching time of $400\,\t{ps}$ for each direction, preceded by an equilibration time of $40\,\t{ps}$. The FL switching was realized for temperatures ranging from $100\K$ to $1300\K$ in intervals of $100\K$. The RS switching for the solid was performed with $T_\t{i} = 100\K$ and $T_\t{f} = 1300\K$, while for the liquid we used $T_{i} = 2000\K$ and  $T_\t{f} = 900\K$. For the UF free-energy calculations we used $p=75$, $\sigma = 1.3\Ang$, and a cutoff radius of $r_\t{c} = 5\sigma$.

\section*{Data availability}
  The data that support the findings of this study are available from the corresponding author upon reasonable request.

\section*{Code availability}
  The code and scripts used to generate the results in this paper can be downloaded from the following repository: \url{https://github.com/freitas-rodrigo/CrystallizationMechanismsFromML}. Any custom code that is not currently available in the repository can be subsequently added to the repository upon request to the corresponding author.


\clearpage
\begin{acknowledgments}
  The authors would like to thank Dr. Vasily Bulatov for providing the script for projecting crystal directions into the standard triangle, Prof. Suneel Kodambaka for bringing to our attention the literature on interface-induced ordering of liquids, and Dr. Pablo Damasceno for the clarifications regarding the entropic origins of layering in weakly-bonded liquids. The authors also acknowledge the fruitful discussions with Prof. Qian Yang, Gowoon Cheon, Evan Antoniuk, and Yanbing Zhu. This work was supported by the Department of Energy National Nuclear Security Administration under Award Number DE--NA0002007, National Science Foundation grants DMREF--1922312 and CAREER--1455050.
\end{acknowledgments}

\section*{Author contributions}
  R.F. performed all calculations and data analysis. R.F. and E.J.R. worked jointly on the interpretation of the data, project design, and manuscript preparation.

\section*{Competing interests}
  The authors declare no competing interests.


\clearpage
\bibliography{bibliography}

\begin{thebibliography}{67}
\providecommand{\natexlab}[1]{#1}
\providecommand{\url}[1]{#1}
\csname url@samestyle\endcsname
\providecommand{\newblock}{\relax}
\providecommand{\bibinfo}[2]{#2}
\providecommand{\BIBentrySTDinterwordspacing}{\spaceskip=0pt\relax}
\providecommand{\BIBentryALTinterwordstretchfactor}{4}
\providecommand{\BIBentryALTinterwordspacing}{\spaceskip=\fontdimen2\font plus
\BIBentryALTinterwordstretchfactor\fontdimen3\font minus
  \fontdimen4\font\relax}
\providecommand{\BIBforeignlanguage}[2]{{%
\expandafter\ifx\csname l@#1\endcsname\relax
\typeout{** WARNING: IEEEtranN.bst: No hyphenation pattern has been}%
\typeout{** loaded for the language `#1'. Using the pattern for}%
\typeout{** the default language instead.}%
\else
\language=\csname l@#1\endcsname
\fi
#2}}
\providecommand{\BIBdecl}{\relax}
\BIBdecl

\bibitem[Asta et~al.(2009)Asta, Beckermann, Karma, Kurz, Napolitano, Plapp,
  Purdy, Rappaz, and Trivedi]{asta_review}
M.~Asta, C.~Beckermann, A.~Karma, W.~Kurz, R.~Napolitano, M.~Plapp, G.~Purdy,
  M.~Rappaz, and R.~Trivedi, ``Solidification microstructures and solid-state
  parallels: Recent developments, future directions,'' \emph{Acta Materialia},
  vol.~57, pp. 941--971, 2009.

\bibitem[Libbrecht(2017)]{libbrecht_review}
K.~G. Libbrecht, ``Physical dynamics of ice crystal growth,'' \emph{Annual
  Review of Materials Research}, vol.~47, pp. 271--295, 2017.

\bibitem[Burton et~al.(1951)Burton, Cabrera, and Frank]{bcf}
W.-K. Burton, N.~Cabrera, and F.~Frank, ``The growth of crystals and the
  equilibrium structure of their surfaces,'' \emph{Philosophical Transactions
  of the Royal Society of London. Series A, Mathematical and Physical
  Sciences}, vol. 243, pp. 299--358, 1951.

\bibitem[Chernov(2012)]{chernov_book}
A.~A. Chernov, \emph{Modern crystallography {III}: crystal growth}.\hskip 1em
  plus 0.5em minus 0.4em\relax Springer Science \& Business Media, 2012,
  vol.~36.

\bibitem[Pimpinelli and Villain(1999)]{pimpinelli_book}
A.~Pimpinelli and J.~Villain, \emph{Physics of crystal growth}.\hskip 1em plus
  0.5em minus 0.4em\relax Cambridge University Press, 1999.

\bibitem[Saito(1996)]{saito_book}
Y.~Saito, \emph{Statistical physics of crystal growth}.\hskip 1em plus 0.5em
  minus 0.4em\relax World Scientific, 1996.

\bibitem[Wilson(1900)]{wilson}
H.~W. Wilson, ``On the velocity of solidification and viscosity of super-cooled
  liquids,'' \emph{The London, Edinburgh, and Dublin Philosophical Magazine and
  Journal of Science}, vol.~50, pp. 238--250, 1900.

\bibitem[Frenkel(1946)]{frenkel}
J.~Frenkel, \emph{Kinetic Theory of Liquids}.\hskip 1em plus 0.5em minus
  0.4em\relax Oxford, The Clarendon Press, 1946.

\bibitem[Stolk et~al.(1993)Stolk, Polman, and Sinke]{si_wf_exp}
P.~Stolk, A.~Polman, and W.~Sinke, ``Experimental test of kinetic theories for
  heterogeneous freezing in silicon,'' \emph{Physical Review B}, vol.~47, p.~5,
  1993.

\bibitem[Sun et~al.(2018{\natexlab{a}})Sun, Xu, and
  Harrowell]{harrowell_nat_mat}
G.~Sun, J.~Xu, and P.~Harrowell, ``The mechanism of the ultrafast crystal
  growth of pure metals from their melts,'' \emph{Nature {M}aterials}, vol.~17,
  p. 881, 2018.

\bibitem[Hoyt and Asta(2002)]{hoyt_prb}
J.~Hoyt and M.~Asta, ``Atomistic computation of liquid diffusivity,
  solid-liquid interfacial free energy, and kinetic coefficient in {A}u and
  {A}g,'' \emph{Physical Review B}, vol.~65, p. 214106, 2002.

\bibitem[Freitas et~al.(2017)Freitas, Frolov, and Asta]{steps_prb}
R.~Freitas, T.~Frolov, and M.~Asta, ``Step free energies at faceted solid
  surfaces: Theory and atomistic calculations for steps on the {C}u (111)
  surface,'' \emph{Physical Review B}, vol.~95, p. 155444, 2017.

\bibitem[Cubuk et~al.(2015)Cubuk, Schoenholz, Rieser, Malone, Rottler, Durian,
  Kaxiras, and Liu]{dogus_prl}
E.~D. Cubuk, S.~S. Schoenholz, J.~M. Rieser, B.~D. Malone, J.~Rottler, D.~J.
  Durian, E.~Kaxiras, and A.~J. Liu, ``Identifying structural flow defects in
  disordered solids using machine-learning methods,'' \emph{Physical Review
  Letters}, vol. 114, p. 108001, 2015.

\bibitem[Schoenholz et~al.(2016)Schoenholz, Cubuk, Sussman, Kaxiras, and
  Liu]{softness}
S.~S. Schoenholz, E.~D. Cubuk, D.~M. Sussman, E.~Kaxiras, and A.~J. Liu, ``A
  structural approach to relaxation in glassy liquids,'' \emph{Nature Physics},
  vol.~12, p. 469, 2016.

\bibitem[Sharp et~al.(2018)Sharp, Thomas, Cubuk, Schoenholz, Srolovitz, and
  Liu]{softness_gb}
T.~A. Sharp, S.~L. Thomas, E.~D. Cubuk, S.~S. Schoenholz, D.~J. Srolovitz, and
  A.~J. Liu, ``Machine learning determination of atomic dynamics at grain
  boundaries,'' \emph{Proceedings of the National Academy of Sciences}, vol.
  115, pp. 10\,943--10\,947, 2018.

\bibitem[Behler and Parrinello(2007)]{parrinello_prl}
J.~Behler and M.~Parrinello, ``Generalized neural-network representation of
  high-dimensional potential-energy surfaces,'' \emph{Physical Review Letters},
  vol.~98, p. 146401, 2007.

\bibitem[Cortes and Vapnik(1995)]{svm}
C.~Cortes and V.~Vapnik, ``Support-vector networks,'' \emph{Machine Learning},
  vol.~20, pp. 273--297, 1995.

\bibitem[Pedregosa et~al.(2011)Pedregosa, Varoquaux, Gramfort, Michel, Thirion,
  Grisel, Blondel, Prettenhofer, Weiss, Dubourg, Vanderplas, Passos,
  Cournapeau, Brucher, Perrot, and Duchesnay]{sklearn}
F.~Pedregosa, G.~Varoquaux, A.~Gramfort, V.~Michel, B.~Thirion, O.~Grisel,
  M.~Blondel, P.~Prettenhofer, R.~Weiss, V.~Dubourg, J.~Vanderplas, A.~Passos,
  D.~Cournapeau, M.~Brucher, M.~Perrot, and E.~Duchesnay, ``Scikit-learn:
  Machine learning in {P}ython,'' \emph{Journal of Machine Learning Research},
  vol.~12, pp. 2825--2830, 2011.

\bibitem[Chang and Lin(2011)]{libsvm}
C.-C. Chang and C.-J. Lin, ``{LIBSVM}: A library for support vector machines,''
  \emph{ACM Transactions on Intelligent Systems and Technology}, vol.~2, pp.
  27:1--27:27, 2011, \url{http://www.csie.ntu.edu.tw/~cjlin/libsvm}.

\bibitem[Fitzner et~al.(2019)Fitzner, Sosso, Cox, and
  Michaelides]{water_nucleation}
M.~Fitzner, G.~C. Sosso, S.~J. Cox, and A.~Michaelides, ``Ice is born in
  low-mobility regions of supercooled liquid water,'' \emph{Proceedings of the
  National Academy of Sciences}, vol. 116, pp. 2009--2014, 2019.

\bibitem[Moore and Molinero(2011)]{valeria_nucleation}
E.~B. Moore and V.~Molinero, ``Structural transformation in supercooled water
  controls the crystallization rate of ice,'' \emph{Nature}, vol. 479, p. 506,
  2011.

\bibitem[Fujiwara(2012)]{si_review}
K.~Fujiwara, ``Crystal growth behaviors of silicon during melt growth
  processes,'' \emph{International Journal of Photoenergy}, vol. 2012, 2012.

\bibitem[Oh et~al.(2005)Oh, Kauffmann, Scheu, Kaplan, and
  R{\"u}hle]{iio_structure}
S.~H. Oh, Y.~Kauffmann, C.~Scheu, W.~D. Kaplan, and M.~R{\"u}hle, ``Ordered
  liquid aluminum at the interface with sapphire,'' \emph{Science}, vol. 310,
  pp. 661--663, 2005.

\bibitem[Oh et~al.(2010)Oh, Chisholm, Kauffmann, Kaplan, Luo, R{\"u}hle, and
  Scheu]{iio_kinetics}
S.~H. Oh, M.~F. Chisholm, Y.~Kauffmann, W.~D. Kaplan, W.~Luo, M.~R{\"u}hle, and
  C.~Scheu, ``Oscillatory mass transport in vapor-liquid-solid growth of
  sapphire nanowires,'' \emph{Science}, vol. 330, pp. 489--493, 2010.

\bibitem[Panciera et~al.(2019)Panciera, Tersoff, Gamalski, Reuter, Zakharov,
  Stach, Hofmann, and Ross]{iio_siau}
F.~Panciera, J.~Tersoff, A.~D. Gamalski, M.~C. Reuter, D.~Zakharov, E.~A.
  Stach, S.~Hofmann, and F.~M. Ross, ``Surface crystallization of liquid
  {A}u--{S}i and its impact on catalysis,'' \emph{Advanced Materials}, vol.~31,
  p. 1806544, 2019.

\bibitem[Kaplan and Kauffmann(2006)]{iio_review}
W.~D. Kaplan and Y.~Kauffmann, ``Structural order in liquids induced by
  interfaces with crystals,'' \emph{Annual Review of Materials Research},
  vol.~36, pp. 1--48, 2006.

\bibitem[de~Jong et~al.(2020)de~Jong, Vonk, Bo{\'c}kowski, Grzegory,
  Honkim{\"a}ki, and Vlieg]{iio_1}
A.~de~Jong, V.~Vonk, M.~Bo{\'c}kowski, I.~Grzegory, V.~Honkim{\"a}ki, and
  E.~Vlieg, ``Complex geometric structure of a simple solid-liquid interface:
  Gan (0001)-ga,'' \emph{Physical Review Letters}, vol. 124, p. 086101, 2020.

\bibitem[Huisman et~al.(1997)Huisman, Peters, Zwanenburg, de~Vries, Derry,
  Abernathy, and van~der Veen]{iio_2}
W.~J. Huisman, J.~F. Peters, M.~J. Zwanenburg, S.~A. de~Vries, T.~E. Derry,
  D.~Abernathy, and J.~F. van~der Veen, ``Layering of a liquid metal in contact
  with a hard wall,'' \emph{Nature}, vol. 390, pp. 379--381, 1997.

\bibitem[Reedijk et~al.(2003)Reedijk, Arsic, Hollander, De~Vries, and
  Vlieg]{iio_3}
M.~Reedijk, J.~Arsic, F.~Hollander, S.~De~Vries, and E.~Vlieg, ``Liquid order
  at the interface of kdp crystals with water: Evidence for icelike layers,''
  \emph{Physical review letters}, vol.~90, p. 066103, 2003.

\bibitem[Teich et~al.(2016)Teich, van Anders, Klotsa, Dshemuchadse, and
  Glotzer]{layering_1}
E.~G. Teich, G.~van Anders, D.~Klotsa, J.~Dshemuchadse, and S.~C. Glotzer,
  ``Clusters of polyhedra in spherical confinement,'' \emph{Proceedings of the
  National Academy of Sciences}, vol. 113, pp. E669--E678, 2016.

\bibitem[Spaepen(1975)]{layering_2}
F.~Spaepen, ``A structural model for the solid-liquid interface in monatomic
  systems,'' \emph{Acta Metallurgica}, vol.~23, pp. 729--743, 1975.

\bibitem[Bracht et~al.(2007)Bracht, Silvestri, Sharp, and
  Haller]{cSi_diffusion}
H.~Bracht, H.~Silvestri, I.~Sharp, and E.~Haller, ``Self-and foreign-atom
  diffusion in semiconductor isotope heterostructures. {II}. {E}xperimental
  results for silicon,'' \emph{Physical Review B}, vol.~75, p. 035211, 2007.

\bibitem[Sciortino et~al.(1991)Sciortino, Geiger, and Stanley]{water_mobility}
F.~Sciortino, A.~Geiger, and H.~E. Stanley, ``Effect of defects on molecular
  mobility in liquid water,'' \emph{Nature}, vol. 354, p. 218, 1991.

\bibitem[Sastry and Angell(2003)]{si_ll}
S.~Sastry and C.~A. Angell, ``Liquid--liquid phase transition in supercooled
  silicon,'' \emph{Nature Materials}, vol.~2, p. 739, 2003.

\bibitem[Plimpton(1995)]{lammps}
S.~Plimpton, ``Fast parallel algorithms for short-range molecular dynamics,''
  \emph{Journal of Computational Physics}, vol. 117, pp. 1--19, 1995,
  \url{https://lammps.sandia.gov}.

\bibitem[Stillinger and Weber(1985)]{stillinger_weber}
F.~H. Stillinger and T.~A. Weber, ``Computer simulation of local order in
  condensed phases of silicon,'' \emph{Physical Review B}, vol.~31, p. 5262,
  1985.

\bibitem[Bulatov and Cai(2006)]{bulatov_book}
V.~Bulatov and W.~Cai, \emph{Computer simulations of dislocations}.\hskip 1em
  plus 0.5em minus 0.4em\relax Oxford University Press on Demand, 2006, vol.~3.

\bibitem[Bussi et~al.(2007)Bussi, Donadio, and Parrinello]{bdp}
G.~Bussi, D.~Donadio, and M.~Parrinello, ``Canonical sampling through velocity
  rescaling,'' \emph{The Journal of Chemical Physics}, vol. 126, p. 014101,
  2007.

\bibitem[Martyna et~al.(1994)Martyna, Tobias, and Klein]{barostat_1}
G.~J. Martyna, D.~J. Tobias, and M.~L. Klein, ``Constant pressure molecular
  dynamics algorithms,'' \emph{The Journal of Chemical Physics}, vol. 101, pp.
  4177--4189, 1994.

\bibitem[Parrinello and Rahman(1981)]{barostat_2}
M.~Parrinello and A.~Rahman, ``Polymorphic transitions in single crystals: A
  new molecular dynamics method,'' \emph{Journal of Applied physics}, vol.~52,
  pp. 7182--7190, 1981.

\bibitem[Tuckerman et~al.(2006)Tuckerman, Alejandre, L{\'o}pez-Rend{\'o}n,
  Jochim, and Martyna]{barostat_3}
M.~E. Tuckerman, J.~Alejandre, R.~L{\'o}pez-Rend{\'o}n, A.~L. Jochim, and G.~J.
  Martyna, ``A {L}iouville-operator derived measure-preserving integrator for
  molecular dynamics simulations in the isothermal--isobaric ensemble,''
  \emph{Journal of Physics A: Mathematical and General}, vol.~39, p. 5629,
  2006.

\bibitem[Shinoda et~al.(2004)Shinoda, Shiga, and Mikami]{barostat_4}
W.~Shinoda, M.~Shiga, and M.~Mikami, ``Rapid estimation of elastic constants by
  molecular dynamics simulation under constant stress,'' \emph{Physical Review
  B}, vol.~69, p. 134103, 2004.

\bibitem[Tuckerman(2010)]{tuckerman_book}
M.~Tuckerman, \emph{Statistical mechanics: theory and molecular
  simulation}.\hskip 1em plus 0.5em minus 0.4em\relax Oxford university press,
  2010.

\bibitem[Rein~ten Wolde et~al.(1996)Rein~ten Wolde, Ruiz-Montero, and
  Frenkel]{alpha}
P.~Rein~ten Wolde, M.~J. Ruiz-Montero, and D.~Frenkel, ``Numerical calculation
  of the rate of crystal nucleation in a {L}ennard-{J}ones system at moderate
  undercooling,'' \emph{The Journal of Chemical Physics}, vol. 104, pp.
  9932--9947, 1996.

\bibitem[Larsen et~al.(2016)Larsen, Schmidt, and Schi{\o}tz]{ptm}
P.~M. Larsen, S.~Schmidt, and J.~Schi{\o}tz, ``Robust structural identification
  via polyhedral template matching,'' \emph{Modelling and Simulation in
  Materials Science and Engineering}, vol.~24, p. 055007, 2016.

\bibitem[Stukowski(2009)]{ovito}
A.~Stukowski, ``Visualization and analysis of atomistic simulation data with
  {OVITO}--the {O}pen {V}isualization {T}ool,'' \emph{Modelling and Simulation
  in Materials Science and Engineering}, vol.~18, p. 015012, 2009,
  \url{http://ovito.org}.

\bibitem[Note1()]{Note1}
Unless otherwise noted, all figures were produced using \protect \texttt
  {Matplotlib}\cite {matplotlib}.

\bibitem[Goodfellow et~al.(2016)Goodfellow, Bengio, and Courville]{dl_book}
I.~Goodfellow, Y.~Bengio, and A.~Courville, \emph{Deep Learning}.\hskip 1em
  plus 0.5em minus 0.4em\relax MIT Press, 2016,
  \url{http://www.deeplearningbook.org}.

\bibitem[Stukowski(2014)]{surface_mesh}
A.~Stukowski, ``Computational analysis methods in atomistic modeling of
  crystals,'' \emph{The Journal of The Minerals, Metals \& Materials Society
  (JOM)}, vol.~66, pp. 399--407, 2014.

\bibitem[Frenkel and Ladd(1984)]{fl}
D.~Frenkel and A.~J. Ladd, ``New {M}onte {C}arlo method to compute the free
  energy of arbitrary solids. {A}pplication to the fcc and hcp phases of hard
  spheres,'' \emph{The Journal of Chemical Physics}, vol.~81, pp. 3188--3193,
  1984.

\bibitem[de~Koning and Antonelli(1996)]{fl_2}
M.~de~Koning and A.~Antonelli, ``Einstein crystal as a reference system in free
  energy estimation using adiabatic switching,'' \emph{Physical Review E},
  vol.~53, p. 465, 1996.

\bibitem[Freitas et~al.(2016)Freitas, Asta, and de~Koning]{cms_freitas}
R.~Freitas, M.~Asta, and M.~de~Koning, ``Nonequilibrium free-energy calculation
  of solids using {LAMMPS},'' \emph{Computational Materials Science}, vol. 112,
  pp. 333--341, 2016.

\bibitem[de~Koning et~al.(1999)de~Koning, Antonelli, and Yip]{rs}
M.~de~Koning, A.~Antonelli, and S.~Yip, ``Optimized free-energy evaluation
  using a single reversible-scaling simulation,'' \emph{Physical Review
  Letters}, vol.~83, p. 3973, 1999.

\bibitem[Schneider and Stoll(1978)]{langevin_thermostat}
T.~Schneider and E.~Stoll, ``Molecular-dynamics study of a three-dimensional
  one-component model for distortive phase transitions,'' \emph{Physical Review
  B}, vol.~17, p. 1302, 1978.

\bibitem[Allen and Tildesley(2017)]{comp_sim_liquids}
M.~P. Allen and D.~J. Tildesley, \emph{Computer simulation of liquids}.\hskip
  1em plus 0.5em minus 0.4em\relax Oxford university press, 2017.

\bibitem[Fultz(2014)]{brent_fultz_book}
B.~Fultz, \emph{Phase Transitions in Materials}.\hskip 1em plus 0.5em minus
  0.4em\relax Cambridge University Press, 2014.

\bibitem[Paula~Leite et~al.(2016)Paula~Leite, Freitas, Azevedo, and
  de~Koning]{uf}
R.~Paula~Leite, R.~Freitas, R.~Azevedo, and M.~de~Koning, ``The uhlenbeck-ford
  model: Exact virial coefficients and application as a reference system in
  fluid-phase free-energy calculations,'' \emph{The Journal of Chemical
  Physics}, vol. 145, p. 194101, 2016.

\bibitem[Leite and de~Koning(2019)]{cms_uf}
R.~P. Leite and M.~de~Koning, ``Nonequilibrium free-energy calculations of
  fluids using {LAMMPS},'' \emph{Computational Materials Science}, vol. 159,
  pp. 316--326, 2019.

\bibitem[Daw and Baskes(1984)]{eam}
M.~S. Daw and M.~I. Baskes, ``Embedded-atom method: Derivation and application
  to impurities, surfaces, and other defects in metals,'' \emph{Physical Review
  B}, vol.~29, p. 6443, 1984.

\bibitem[Foiles et~al.(1986)Foiles, Baskes, and Daw]{eam_cu}
S.~Foiles, M.~Baskes, and M.~S. Daw, ``Embedded-atom-method functions for the
  fcc metals cu, ag, au, ni, pd, pt, and their alloys,'' \emph{Physical review
  B}, vol.~33, p. 7983, 1986.

\bibitem[Sun et~al.(2018{\natexlab{b}})Sun, Xu, and Harrowell]{ultrafast}
G.~Sun, J.~Xu, and P.~Harrowell, ``The mechanism of the ultrafast crystal
  growth of pure metals from their melts,'' \emph{Nature materials}, vol.~17,
  p. 881, 2018.

\bibitem[Hunter(2007)]{matplotlib}
J.~D. Hunter, ``Matplotlib: A 2{D} graphics environment,'' \emph{Computing in
  Science \& Engineering}, vol.~9, pp. 90--95, 2007.

\bibitem[Stukowski and Albe(2010)]{DXA_1}
A.~Stukowski and K.~Albe, ``Extracting dislocations and non-dislocation crystal
  defects from atomistic simulation data,'' \emph{Modelling and Simulation in
  Materials Science and Engineering}, vol.~18, p. 085001, 2010.

\bibitem[Stukowski et~al.(2012)Stukowski, Bulatov, and Arsenlis]{dxa}
A.~Stukowski, V.~V. Bulatov, and A.~Arsenlis, ``Automated identification and
  indexing of dislocations in crystal interfaces,'' \emph{Modelling and
  Simulation in Materials Science and Engineering}, vol.~20, p. 085007, 2012.

\bibitem[Galvin et~al.(1985)Galvin, Mayer, and Peercy]{si_epitaxial}
G.~Galvin, J.~Mayer, and P.~Peercy, ``Solidification kinetics of pulsed laser
  melted silicon based on thermodynamic considerations,'' \emph{Applied Physics
  Letters}, vol.~46, pp. 644--646, 1985.

\bibitem[Hedler et~al.(2004)Hedler, Klaum{\"u}nzer, and Wesch]{si_glass}
A.~Hedler, S.~L. Klaum{\"u}nzer, and W.~Wesch, ``Amorphous silicon exhibits a
  glass transition,'' \emph{Nature Materials}, vol.~3, p. 804, 2004.

\bibitem[Shou and Pan(2016)]{gamma_si}
W.~Shou and H.~Pan, ``Silicon-wall interfacial free energy via thermodynamics
  integration,'' \emph{The Journal of chemical physics}, vol. 145, p. 184702,
  2016.

\end{thebibliography}


\clearpage

\end{document}